\documentclass[traditabstract]{aa}
\usepackage{txfonts,textcomp,natbib,textcomp,amssymb}
\usepackage{graphicx}
\bibpunct{(}{)}{;}{a}{}{,} 

\begin{document}

\title{Heating and cooling of the neutral ISM in the NGC\,4736 circumnuclear ring}
\titlerunning{Heating-cooling NGC\,4736 circumnuclear ring}

\author{T.\,P.\,R. van der Laan \inst{1,2}
\and L. Armus \inst{2}
\and P. Beirao \inst{3}
\and K. Sandstrom \inst{4}
\and B. Groves \inst{5}
\and E. Schinnerer \inst{5}
\and B.T. Draine \inst{6}
\and J.D. Smith \inst{7}
\and M. Galametz \inst{8}
\and M. Wolfire \inst{9}
\and K. Croxall \inst{10}
\and D. Dale \inst{11}
\and R. Herrera Camus \inst{9}
\and D. Calzetti \inst{12}
\and R.C. Kennicutt, Jr. \inst{13}
}

\institute{Institute de Radioastronomie Millimetrique (IRAM), 300 Rue de la Piscine, 38406 St. Martin d'Heres, Grenoble, France; \\ \textit{vanderlaan@iram.fr}
\and Spitzer Science Center, California Institute of Technology, MC 314-6, Pasadena, CA 91125, USA
\and Observatoire de Paris, 61 avenue de l'Observatoire, F-75014 Paris, France
\and Steward Observatory, University of Arizona, 933 N. Cherry Ave, Tucson, AZ 85721, USA
\and Max-Planck-Institut f\"ur Astronomie, K\"onigstuhl 17, 69117 Heidelberg, Germany
\and Department of Astrophysical Sciences, Princeton University, Princeton, NJ 08544, USA
\and Department of Physics and Astronomy, University of Toledo, Toledo, OH 43606, USA
\and European Southern Observatory, Karl-Schwarzchild-Str. 2, D-85748 Garching-bei-M\"{u}nchen, Germany
\and Department of Astronomy, University of Maryland, College Park, MD 20742, USA
\and Department of Astronomy, The Ohio State University, 140 West 18th Avenue, Columbus, OH 43210, USA
\and Department of Physics \& Astronomy, University of Wyoming, Laramie, WY 82071, USA
\and Department of Astronomy, University of Massachusetts, Amherst, MA 01003, USA
\and Institute of Astronomy, University of Cambridge, Madingley Road, Cambridge CB3 0HA, UK
}

\abstract{The manner in which gas accretes and orbits within circumnuclear rings has direct implications for the star formation process. In particular, gas may be compressed and shocked at the inflow points, resulting in bursts of star formation at these locations. Afterwards the gas and young stars move together through the ring.  In addition, star formation may occur throughout the ring, if and when the gas reaches sufficient density to collapse under gravity.  These two scenarios for star formation in rings are often referred to as the `pearls on a string' and `popcorn' paradigms. In this paper, we use new Herschel PACS observations, obtained as part of the KINGFISH Open Time Key Program, along with archival Spitzer and ground-based observations from the SINGS Legacy project, to investigate the heating and cooling of the interstellar medium in the nearby star-forming ring galaxy, NGC\,4736.  By comparing spatially resolved estimates of the stellar FUV flux available for heating, with the gas and dust cooling derived from the FIR continuum and line emission, we show that while star formation is indeed dominant at the inflow points in NGC 4736, additional star formation is needed to balance the gas heating and cooling throughout the ring. This additional component most likely arises from the general increase in gas density in the ring over its lifetime.  Our data provide strong evidence, therefore, for a combination of the two paradigms for star formation in the ring in NGC\ 4736.}


\maketitle 

\section{Introduction}
Circumnuclear star formation rings are present in approximately 20\% of nearby disk galaxies \citep{Comeron2010}. Early theories suggested that the radius at which the rings form was the inner Lindblad resonance (ILR) radius, which is set by an asymmetric component in the stellar disk such as a large scale bar. However, further modeling of galaxy gravitational potentials has shown that circumnuclear ring formation is linked more particularly to the presence of a sufficient number of $x_2$ orbits \citep{Regan2003}. Such orbits are (near-)circular, in contrast to the elongated $x_1$ orbits at larger radii, which make up the large scale bar \citep{1980A&A....92...33C}. The radii of $x_2$ orbits are close to the ILR radius.

As gas and dust lose angular momentum, they move inward. At the overlap between $x_1$ to $x_2$ orbits, gas and dust will settle on the lower energy $x_2$ orbits as a result of collisions, forming a circumnuclear ring. The discrete locations where this transition happens are generally called `inflow points'. Such inflow points are most obvious in ring systems with large scale stellar bars, where the connection between the spiral shocks of the bar and the circumnuclear ring can often be clearly seen \citep[for example NGC\,6951,][]{Tessel}. However, circumnuclear rings also exist in galaxies classified as non-barred \citep[see, e.g.,][for a discussion on the difference, or lack thereof, between barred and non-barred systems in this context]{2010AJ....139.2465G}. Any gravitational asymmetry in the disk may suffice, as long as enough $x_2$ orbits are present in the gravitational potential.

The gas and dust can experience shocks as they transition from $x_1$  to $x_2$  orbits at the inflow points. However, after the gas and dust settles on the $x_2$ orbits of the ring, the accreted ISM experiences very little large-scale friction. With that in mind, it is not too surprising that it has been found in multiple galaxies that most (if not all) star formation in circumnuclear rings takes place at the inflow points \citep{2001MNRAS.323..663R,2005ApJ...633L..25A,Boker2008,2008ApJS..174..337M,Tessel2,Tessel3,2014MNRAS.438..329F}. Consequently, it is thought that, owing to the absence of large scale friction or temporal variations in gravitational potential, stars and gas primarily move through the circumnuclear ring together. The stellar age gradient regularly seen stretching out from each inflow point is often referred to as `pearls-on-a-string', first coined by \citet{Boker2008}. An alternative theory \citep{Elmegreen1994} suggests that the inflow points play no special role in star formation, and that a build up of gas will trigger a burst of star formation after a critical limit has been reached anywhere in a ring. This is often referred to as the `popcorn' theory of star formation in rings.

Recent star formation is often tracked by observing hydrogen recombination lines such as H$\alpha$ or Br$\gamma$. The equivalent width of the Balmer lines is, for example, dependent on stellar age from $\sim$4\,Myr \citep[e.g.][]{1995ApJS...96....9L,1997AJ....113..975B,1999ApJS..125..489G}. Alternatively, emission lines from other atoms and molecules stimulated by stellar radiation, such as warm H$_2$, HeI, and [FeII], in the optical and NIR can be used. With a combination of these lines it is also possible to observe potential age gradients, as demonstrated for the circumnuclear ring in NGC\,613  \citep{Boker2008,2014MNRAS.438..329F}. However, strong extinction can make observations in the optical and NIR range difficult, especially at the inflow points where dust and gas build up, potentially limiting the effectiveness of these methods.

When there are no large-scale mechanical stresses, stellar radiation is most likely to be the prevalent form of heating for the neutral gas in the ring (excluding contributions of CR or X-ray ionization). Thus, an alternative way of tracking recent star formation is to study the heating and cooling of the neutral interstellar medium (ISM) via strong emission lines of the neutral gas in the FIR. In the presence of star formation, the heating of the neutral ISM is dominated by photo-electric heating \citep{1994ApJ...427..822B,2001ApJS..134..263W,2001ApJ...548L..73H}. Dust grains and PAHs eject electrons as the result of the absorption of FUV (6eV $<$ h$\nu$ $<$ 13.6eV) photons from young O and B type stars. These electrons in turn interact with the gas, heating atoms and molecules. The cooling times for dust and gas are much shorter than the Myr astronomical timescales of interest here. Thus, gas heating and cooling, via photo-electric heating, may be considered in balance at all times, and the observed cooling flux can be used as a proxy for the heating.

To test whether the heating of the neutral ISM in the ring is indeed dominated by stellar radiative heating from stars formed at the inflow points, we investigate the case of the circumnuclear ring in NGC\,4736, a nearby (D=4.66\,Mpc, $i\sim35\degr$, PA$\sim300\degr$) SAab spiral galaxy. The ring in NGC\, 4736 has a radius of 40$\arcsec$ (0.9\,kpc). Based on a previously observed HI velocity field \citep{2008AJ....136.2563W} and assuming that the spiral arms are concave, the north side of the disk is the near side, and the south the far side, with a clockwise rotation. We have combined available data to form a large multiwavelength dataset, tracing the gas and the recent star formation, including {\it Herschel} PACS data from the KINGFISH (Key Insights on Nearby Galaxies: A Far-Infrared Survey with Herschel) Open Time Key Program \citep{2011PASP..123.1347K}. The PACS observations enable us to probe the main neutral gas cooling lines of $[$CII$]$ at 158$\mu$m and $[$OI$]$ at 63$\mu$m, as well as the FIR dust continuum. We also include data obtained with the {\it Spitzer} Space Telescope as part of the SINGS ({\it Spitzer} Infrared Nearby Galaxies Survey) Legacy program \citep{2003PASP..115..928K}, to probe the mid-infrared wavelength range, which includes emission from large dust grains, PAH's, stars, and molecular and atomic gas.

In \S2 we present the data. In \S3 the atomic and molecular gas mass surface densities, and subsequent dust mass surface density, are computed. In \S4 a census is made of the star formation rate density throughout the ring. In \S5 the contribution of the neutral ISM to the observed flux is determined. We then compare the dust and gas cooling with the stellar heating in \S6. The implications are discussed in \S7, with a summary in \S8.

\section{Data sets}
\begin{figure}
\resizebox{\hsize}{!}{\includegraphics[width=17cm]{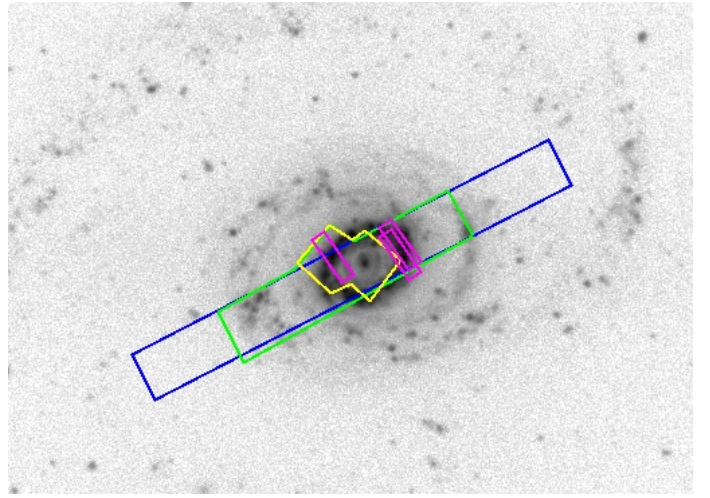}}
\caption{Spatial coverage of the spectral data sets presented in this study, overlaid on the {\it GALEX} FUV image of NGC\,4736. The circumnuclear ring at a radius of r$\sim$40$\arcsec$ (0.9\,kpc) stands out clearly in the {\it GALEX} image. The different regions mapped spectroscopically, and indicated here in color, are: {\it Spitzer/IRS} long-low (14-38 $\mu$m) spectra (dark blue contours), {\it Herschel/PACS} [CII] and [OI] emission line maps (green contours), {\it Herschel/PACS} [NII]122$\mu$m, [NII]205$\mu$m emission line maps (yellow contours), and {\it Spitzer/IRS} short-low (5-14$\mu$m) spectra (purple contours).  In this image, north is up and east is to the left, and the total size is approximately 7.3 arc minutes on a side.}
\label{fig:overlay}
\end{figure}

\subsection{Herschel PACS}\label{sect:data}
Both spectra and images were obtained for NGC\,4736 with the {\it Herschel} PACS instruments as part of the KINGFISH Open Time Key Program. The PACS spectroscopic instrument is an IFU (integral field unit) with 5$\times$5, 9\farcs4  pixels \citep{2010A&A...518L...2P}. This gives an instantaneous Field-of-View (FoV) of 47\arcsec\,$\times$\,47\arcsec. The PACS beam FWHM is 5\farcs2 in the wavelength range 60-85\,$\mu$m, and 12\arcsec\, at 130-210\,$\mu$m. For the spectra several targeted pointings were made for the various lines. The [NII]122$\mu$m and [NII]205$\mu$m lines were observed with 2 adjacent pointings, one focussed on the nucleus and one on the circumnuclear ring (see yellow contours in Fig. \ref{fig:overlay}). The [CII]158$\mu$m and [OI]63$\mu$m lines were observed in a strip 200\arcsec\, long (green contours), with an orientation equal to previous {\it Spitzer} IRS data (blue contours), and one adjacent pointing on the circumnuclear ring (equal to the [NII] off-nucleus pointing). The FIR continuum was probed at 70\,$\mu$m, 100\,$\mu$m, and 160\,$\mu$m with the PACS photometer over the full extent of the galaxy. 

The PACS observations were reduced consistently for the complete KINGFISH sample using the Scanamorphus package \citep{2013PASP..125.1126R}. The KINGFISH observations are described in \citet{2012ApJ...745...95D} and  \citet{2012ApJ...747...81C}.  In all cases, the images were rescaled to a pixel scale of 4\arcsec. The spectra zeroth-moment maps were convolved to the PACS\,160$\mu$m resolution of 12\arcsec\,, using a pixel scale of 4\arcsec\,, since the PACS\,160 beam is essentially the same as the [CII] beam.

\begin{figure*}
\resizebox{\hsize}{!}{\includegraphics[width=17cm]{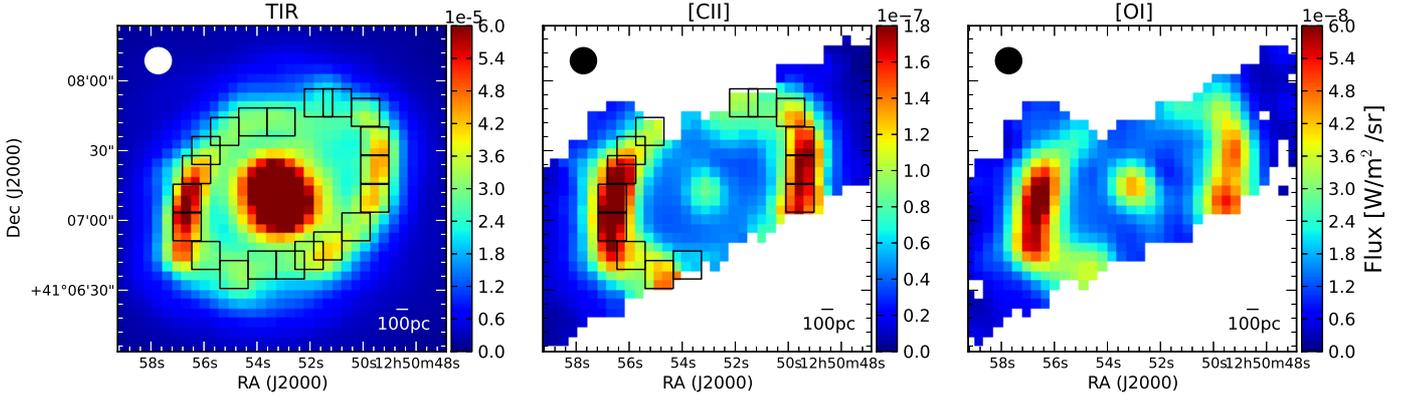}}
\caption{Distribution of the Total InfraRed (TIR) continuum emission ({\it left}), [CII]\,158\,$\mu$m line emission ({\it middle}), and [OI]\,63\,$\mu$m line emission ({\it right}) in the circumnuclear ring of NGC\ 4736. The 12\arcsec\, apertures used to extract fluxes are indicated in the TIR and [CII] panels. The spatial scale is given in the bottom right corner, and the PACS\,160$\mu$m beam of $\sim$12\arcsec\, is indicated in the top left corner of each panel.}
\label{fig:FIRgas}
\end{figure*}

The [CII] and [OI] intensity maps, as well as the total-IR (TIR) continuum map are shown in Fig. \ref{fig:FIRgas}. The TIR continuum is constructed from a combination of four individual Spitzer and Herschel images, at 8\,$\mu$m, 24\,$\mu$m, 70\,$\mu$m, and 160\,$\mu$m, following the equation in \citet{2007ApJ...657..810D},

\begin{equation}\label{eq1}
TIR = \left(0.95\langle\nu\,S_{\nu}\rangle_{8} + 1.15\langle\nu\,S_{\nu}\rangle_{24} + \langle\nu\,S_{\nu}\rangle_{70} + \langle\nu\,S_{\nu}\rangle_{160}\right).
\end{equation}

\noindent Even though the FoV of the spectra clips the northern and southern part of the circumnuclear ring, the [CII] and [OI] emission is clearly concentrated there. The TIR continuum also has elevated emission at those locations, but is mostly dominated by the nucleus. 
 
Integrated fluxes are extracted by covering the ring with 3x3 pixel (12\arcsec\,x12\arcsec\,) apertures. In this manner we measured the [CII] and [OI] emission line and continuum fluxes at 13 individual points, seven on the east side, and six on the west side of the ring. For the TIR continuum an additional six apertures are positioned on the ring outside the spectroscopic FoV (see Fig. \ref{fig:FIRgas}). This east (west) side corresponds to the blue (red) shifted side of the galaxy. The positions of, and fluxes in, these 12\arcsec\, apertures are given in Table \ref{tab:FIRfluxes}.

\begin{table*}
\begin{minipage}{2\columnwidth}
\centering
\caption{[CII], [OI], and TIR surface brightnesses in defined apertures}\label{tab:FIRfluxes}
\begin{tabular}{r c c c c c c c c c}
\hline\hline
\# & pos x & pos y & Angle & [CII] & [CII]$_{unc}$ & [OI] & [OI]$_{unc}$ & TIR & TIR$_{unc}$ \\
 & [pix] & [pix] & [deg] & W/m$^2$/sr & rel. & W/m$^2$/sr & rel. & W/m$^2$/sr & rel. \\
 \hline
 1 &149  &144 & 15 & -- & -- & -- & -- & 1.29e-4 & 0.09 \\
 2  &146  &143 & 29 & 1.12e-6  & 0.01  & 2.43e-7  & 0.12  & 2.64e-4  & 0.07 \\
 3  &143  &145 & 49 & 1.13e-6  & 0.01  & 3.33e-7  & 0.06  & 2.84e-4  & 0.04 \\
 4  &141  &148 & 68 & 1.62e-6  & 0.01  & 4.91e-7  & 0.04  & 4.54e-4  & 0.04 \\
 5  &141  &151 & 85 & 1.75e-6  & 0.01  & 5.40e-7  & 0.03  & 4.88e-4  & 0.03 \\
 6  &142  &154 & 102 & 1.60e-6  & 0.01  & 4.24e-7  & 0.05  & 3.98e-4  & 0.03 \\
 7  &143  &156 & 117 & 1.11e-6  & 0.01  & 2.72e-7  & 0.09  & 2.94e-4  & 0.03 \\
 8  &145  &158 & 135 & 8.87e-7  & 0.02  & 2.00e-7  & 0.23  & 2.56e-4  & 0.04 \\
 9  &148  &159 & 157 & -- & -- & -- & -- & 2.70e-4 & 0.03 \\
10 &151  &159 & 180 & -- & -- & -- & -- & 2.56e-4 & 0.03 \\
11 &155  &161 & 205 & 8.50e-7  & 0.02  & 1.31e-7  & 0.19  & 2.20e-4  & 0.03 \\
12 &157  &161 & 213 & 7.83e-7  & 0.02  & 1.69e-7  & 0.11  & 2.12e-4  & 0.03 \\
13 &160  &160 & 229 & 1.05e-6  & 0.01  & 2.94e-7  & 0.06  & 2.66e-4  & 0.03 \\
14 &161  &157 & 243 & 1.34e-6  & 0.01  & 3.70e-7  & 0.05  & 3.28e-4  & 0.03 \\
15 &161  &154 & 259 & 1.44e-6  & 0.01  & 3.82e-7  & 0.05  & 3.54e-4  & 0.03 \\
16 &161  &151 & 275 & 1.40e-6  & 0.01  & 4.34e-7  & 0.08  & 3.38e-4  & 0.03 \\
17 &159  &148 & 297 & -- & -- & -- & -- & 2.88e-4 & 0.05 \\
18 &156  &146 & 320 & -- & -- & -- & -- & 3.42e-4 & 0.10 \\
19 &154  &145 & 337 & -- & -- & -- & -- & 3.14e-4 & 0.12 \\
20 &152  &144 & 353 & -- & -- & -- & -- & 2.56e-4 & 0.12\\
\hline
\end{tabular}\\
\begin{flushleft}
{\bf Notes:} [CII] and [OI] emission line fluxes, TIR continuum flux, and their relative uncertainties in 3x3 pixel (12\arcsec\,x12\arcsec\,) apertures. Central positions (in TIR map pixels) of the apertures are given in the 2nd and 3rd columns. The order of the apertures (angles) here is clockwise (subsequently following the direction of gas motion), starting from due south. For reference, pixel (151,152) corresponds to the galaxy center (RA 12h50m53.18s, Dec 41d07m12.0s).
\end{flushleft}
\end{minipage}
\end{table*}

\subsection{Spitzer}
As part of the SINGS survey NGC\,4736 was observed with {\it Spitzer}-IRS using the `short-low' and `long-low' spectral modules in spectral mapping mode. The circumnuclear ring is incompletely covered by the `long-low' footprint (see again Fig. \ref{fig:overlay}, blue contour), and partially by the three off-nuclear `short-low' pointings (purple contours). The `short-low' (SL) module has a pixel scale of 1\farcs85 and a resolving power of 60-120, decreasing with increasing wavelength. The observations have a FoV of 3\farcs6 by 27\arcsec\, and cover the wavelength range 5-14$\mu$m. The `long-low' (LL) module has a pixel scale of 5\farcs1, a resolving power of approximately 60-120, again decreasing with wavelength, a FoV of 10\farcs6 by 84\arcsec\,, and covers the wavelength range 14-28$\mu$m. The reduced and fully calibrated data cubes are available on the SINGS Spitzer and Ancillary Data website\footnote{http://irsa.ipac.caltech.edu/data/SPITZER/SINGS/}.

For this work we extracted the [SIII] emission line flux at 18.7\,$\mu$m and 33.5\,$\mu$m by fitting the lines and the underlying warm dust continuum with PAHFIT \citep{2007ApJ...656..770S} on the LL data cube alone. The [SIII] emission lines are not blended, and the dust emission is a smoothly varying composition of black-body curves of different temperatures. Consequently, the separation of line and continuum emission at these wavelengths with PAHFIT is straight-forward. A check at the positions with both SL and LL data showed that the exclusion of the SL data cubes had no detrimental effect on the fitting of the long-wavelength dust continuum. Since the IRS data are not combined with the other data sets - they will only be used in Sect. \ref{sect:gn} - and have a limited FoV, no effort was made to convolve the cubes to the PACS160 resolution. 

Spitzer-IRAC imaging at 8\,$\mu$m and 24\,$\mu$m are also presented, and were convolved to the PACS160 resolution using kernels from \citet{2011PASP..123.1218A}.

\subsection{Ancillary data}
The THINGS survey\footnote{http://www.mpia-hd.mpg.de/THINGS/Overview.html} \citep{2008AJ....136.2563W} observed the HI distribution of NGC\,4736 with the VLA between March 2003 and March 2004 in the BCD configurations. The HI intensity maps (natural and robust weighting) are publicly available from the THINGS website. The beam size of the observations is, depending on the chosen weighting, 10\farcs22\,$\times$\,9\farcs07\, (natural), or 5\farcs96\,$\times$\,5\farcs55\, (robust).

\begin{figure*} 
\resizebox{\hsize}{!}{\includegraphics[width=17cm]{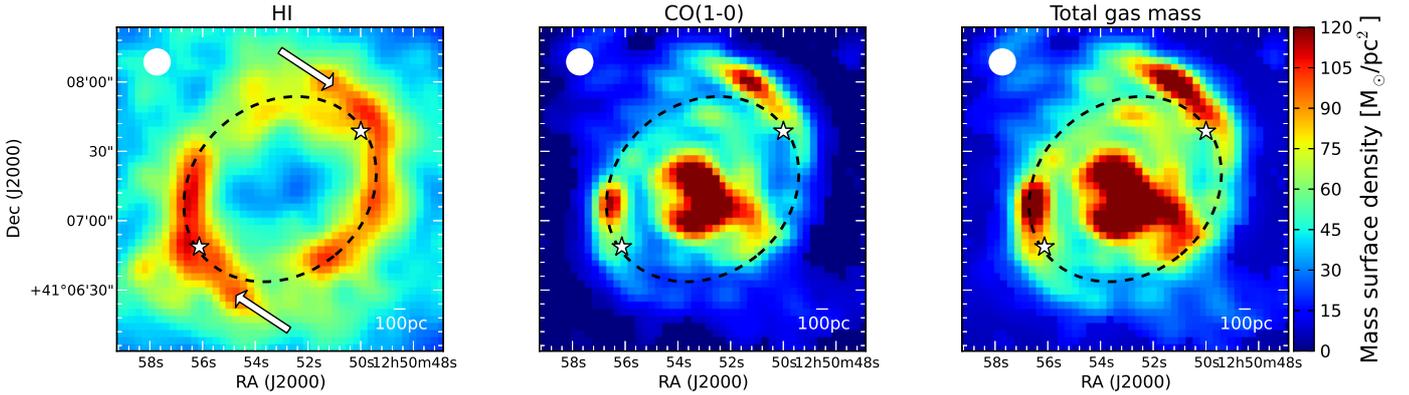}}
\caption{{\it Left and middle:} Mass surface density maps of atomic (HI) and molecular (CO(1-0)) gas, at the PACS\,160$\mu$m ($\sim$12\arcsec) resolution. {\it Right:} Total gas mass surface density map, based on HI and CO maps, with a 26\% correction for helium. The radius of the circumnuclear ring and the locations of the inflow points are indicated with a dashed circle and stars, respectively. Clockwise gas flow is indicated with the white arrows.}
\label{fig:gas}
\end{figure*}

The BIMA SONG survey \citep{2003ApJS..145..259H} observed the CO(1-0) distribution in NGC\,4736 in October 1997. The total FoV of these data are $\sim$190\arcsec\, (4.3\,kpc). The data have a resolution of 6\arcsec\, and are short spacing corrected. The resulting data cube and moment maps are publicly available from NED. 

NGC\,4736 was observed with GALEX in both the NUV and FUV bands between April and June, 2004. The observations are publicly available through the MAST GALEX public access site\footnote{http://galex.stsci.edu/GR6/}. The FoV is 1.2\degr\,, with a FWHM angular resolution of 4\farcs0 (NUV) and 5\farcs6 (FUV).

An H$\alpha$ map was obtained from the NASA/IPAC Extraglactic Database (NED), but was originally observed, reduced, and continuum and [NII] subtracted by \citet{2004A&A...426.1135K}. It was obtained with the Isaac Newton 2.5m telescope. The angular resolution is 1\farcs4$\times$1\farcs4 and the pixel scale is 0\farcs3.

\section{ISM distribution in the ring}
\subsection{Atomic and molecular gas}
Both the HI and CO(1-0) intensity distributions, convolved to the PACS\,160$\mu$m resolution, are shown in the first two panels of Fig. \ref{fig:gas}. The HI distribution is dominated by two arcs which are co-spatial with the circumnuclear star forming ring (whose radius is approximated with the dashed ellipse, representing an $i\sim35\degr$ inclined ring). The nucleus is devoid of atomic gas. By comparison, the CO(1-0) distribution shows emission in a more spiral arm-like pattern. (This is even more apparent at native resolution.) The spiral arms only partly overlap with the circumnuclear ring. Unlike the HI, the CO(1-0) intensity distribution further shows strong emission from the nucleus.

The `total' gas mass density is computed via these two tracers, by assuming a CO-conversion factor, X$_{CO}$, of 7.4\,$\times$\,$10^{19}$ [cm$^{-2}$\,(K km/s)$^{-1}$] \citep{2013ApJ...777....5S}. This value is lower than the Galactic X$_{CO}$ value of 2.0\,$\times$\,$10^{20}$ [cm$^{-2}$\,(K km/s)$^{-1}$] most often assumed for nearby galaxies, but \citet{2013ApJ...777....5S} have shown that X$_{CO}$ drops to this value towards the galactic nuclear region of this galaxy (averaged at a resolution of 37\farcs5). Due to the lower X$_{CO}$ factor, the molecular gas composes a smaller fraction of the total gas mass. The total gas mass map is still dominated by molecular gas, as can be expected in the molecule rich centers of galaxies, but atomic gas provides about 1/3 of the gas mass in the ring. A 26\% correction for helium and metals has been applied to the total gas map. 

Ring inflow points are defined as the position where the gas spiral arm(s) cross onto the ring. In a bi-symmetric potential, e.g. with a stellar bar, the two points should be more or less symmetric about the nucleus. An indication of their position in this galaxy is given in Fig. \ref{fig:gas}. The total gas mass map shows two, possibly three, regions of high mass surface density at the radius of the ring, in the east, the northwest, and at lower significance to the southwest. The first two are near the inflow points.

\subsection{Dust}
By comparing our gas mass map with a map of the dust mass, we can determine whether or not the dust and gas in this region are well mixed, which is a necessary condition for our upcoming analysis and comparison to models of the star formation in the ring. The dust mass distribution in NGC\,4736 is derived using via the \citet{2007ApJ...657..810D} model, including Spitzer, Herschel PACS\footnote{the same observations as used in this work} and SPIRE observations of the source, to constrain the IR dust emission  \citep{2012ApJ...756..138A}. We find that our mass distribution of gas shows a one-on-one morphology correspondence, confirming that dust and gas are qualitatively well mixed. Equally, when we compare our TIR continuum map (Fig. \ref{fig:FIRgas}) with the dust luminosity found by the model a one-on-one morphology correspondence can be observed. Additionally we find that our total gas mass map also matches quantitatively, i.e., with reasonable conversion values for the dust-to-gas ratio (DGR) we recover the values in the dust map.

\begin{figure*}
\resizebox{\hsize}{!}{\includegraphics[width=17cm]{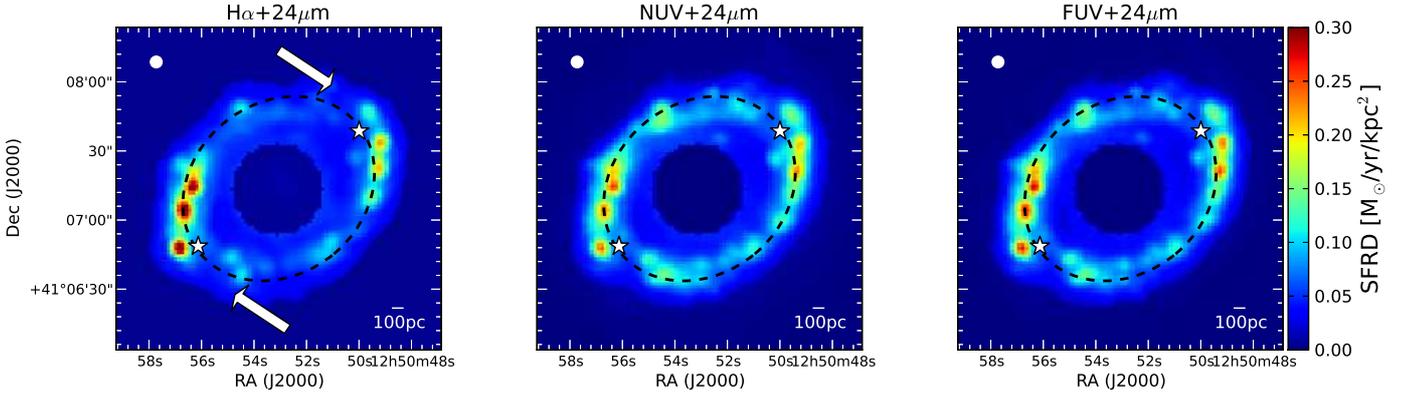}}
\caption{Star formation rate density maps, based on H$\alpha$+24$\mu$m ({\it left}), NUV+24$\mu$m ({\it middle}), and FUV+24$\mu$m ({\it right}), at the Spitzer 24$\mu$m ($\sim$5\arcsec) resolution. The H$\alpha$+24$\mu$m derived SFRD values in this image have been multiplied by a factor two for visualization purposes. The galactic nucleus and larger radii have been masked. The radius of the circumnuclear ring and the locations of the inflow points are indicated with a dashed circle and stars, respectively. The Spitzer 24$\mu$m beam of $\sim$5\arcsec\, is indicated in the top left corner of each subpanel.}
\label{fig:SFR}
\end{figure*}

\section{Star formation distribution in the ring}\label{sect:SFR}
The star formation rate in the circumnuclear ring can be derived from either the H$\alpha$, NUV, or FUV data sets. To compensate for the stellar light absorbed by dust, all three star formation rate tracers are combined with the Spitzer 24$\mu$m image to obtain dust-corrected values. 

The star formation rate is computed based on the SFR calibrations given in \citet{2011ApJ...737...67M}, \citet{2011ApJ...741..124H}, and \citet{2012ARA&A..50..531K}. The resulting SFR surface density (SFRD) maps are shown in Fig. \ref{fig:SFR}. Differences between the three maps reflect the uncertainties and physical/temporal variations among the SFR tracers. The region for which the SFRD is computed is, in all cases, limited at small radii by a circular mask at 20\arcsec\, to exclude emission from the nucleus. At the outer edge, a limit has been placed at the minimum of either 70\arcsec\, radius, or the position where the flux in the tracers falls to a 2$\sigma$ level. The maps shown in Fig. \ref{fig:SFR} are at the resolution of the 24$\mu$m observations, which is $\sim$5\arcsec\,, and have a pixel scale of 1\farcs5. SFRD maps at the PACS\,160\,$\mu$m resolution were also computed.

In all three panels, the regions of highest star formation rates are found in the southeast and northwest quadrants of the ring. These positions are near/at the ring inflow points, as expected. At the other two quadrants of the ring, the SFRD is lower (by factors of three to six). This could indicate a secondary star formation event of lower-level intensity, or the continued emission from an ageing burst. The SFRD contrast around the ring is strongest in the H$\alpha$+24$\mu$m map. The SFRD values found for H$\alpha$+24$\mu$m were checked against previous results by \citet{2004ApJ...605..183W}. \citet{2004ApJ...605..183W} found a SFRD for the circumnuclear ring, based on H$\alpha$ fluxes alone, of 0.04\,M$_{\sun}$/yr/kpc$^2$. We match this value with our H$\alpha$ observations. 

Comparing with the gas maps shown in Fig. \ref{fig:gas}, the SFR distribution is most similar to that of the atomic gas. This is surprising, given the fact that stars form from molecular not atomic gas. We speculate here that the young stars in the ring of NGC\ 4736 are energetic enough to dissociate large quantities of molecular gas in their immediate vicinity (without ionizing that gas as well), and will investigate this matter in an upcoming paper (Van der Laan, in prep.)

The star formation regions near the inflow points each cover about a fifth of the ring. The orbital time for a star at this radius (r=0.9\,kpc) is about 25\,Myr, given a 220\,km/s rotational velocity (value derived from the BIMA SONG first moment map). Therefore the duration of the starburst cannot be more than 5\,Myr.  This is also a reasonable time span to explain the difference between the H$\alpha$+24$\mu$m, and NUV+24$\mu$m and FUV+24$\mu$m maps. When we `subtract' the H$\alpha$ distribution from either the NUV or FUV distribution (not shown), what remains is a more uniform distribution of flux, expected if the UV emission drops more slowly with time ($\sim$5.5\,Myr for H$\alpha$, $\sim$11\,Myr for NUV/FUV, as based on Starburst99 stellar burst synthesis models). The UV tracers are still tracing the previous `event' when they have moved through half the ring.

With these observations it is not possible to directly distinguish an age-gradient of star formation in the hot spots seen at the inflow points. Inspection of the observations shows that the hot spots are strongest in the 24\,$\mu$m observations and nearly unseen in the FUV and H$\alpha$, implying they are (strongly) extincted regions. However, an equivalent width map of the H$\alpha$ emission, given that a suitable continuum image is found, could potentially show age-related variations between the hot spots.

Averaged over the ring, the SFRD values correspond to a star formation rate density of $\sim$0.065\,M$_{\sun}$/yr/kpc$^2$. Within individual hotspots, the SFRD reaches more than 0.25\,M$_{\sun}$/yr/kpc$^2$. High SFRD are a common characteristic of circumnuclear star forming rings  \citep{2004ARA&A..42..603K,2012ARA&A..50..531K}. Integrating over the full area of the ring leads to SFR values of 0.16, 0.27, and 0.25\,M$_{\sun}$/yr, for H$\alpha$+24$\mu$m, NUV+24$\mu$m, and FUV+24$\mu$m, respectively. These values are lower than previously derived SFRs for the whole galaxy of 0.38\,M$_{\sun}$/yr \citep{2010ApJ...714.1256C}, 0.43\,M$_{\sun}$/yr \citep{2006PhDT........12L}, and 0.70\,M$_{\sun}$/yr \citep{2011ApJ...738...89S}, but show the major role the circumnuclear ring plays in the overall star formation in NGC\ 4736. 

\section{Radiation field in the ring}\label{sect:gn}
In order to investigate the effect the young stars have on the local ISM, it is important to measure the emission from the dense neutral regions (the PDRs) close to the forming stars.

As a first step, it is important to estimate, and then remove, the contribution of [CII] from the diffuse, ionized ISM. Neutral Carbon has an ionization potential (11.3eV) which is slightly lower than that of Hydrogen (13.6eV). Consequently, it will be present in both the neutral and ionized ISM. The contribution of [CII] from the diffuse ionized ISM can be determined from an observed [CII]/[NII]205\,$\mu$m ratio, since N$^+$ emission arrises exclusively from the diffuse ionized medium \citep{1994ApJ...434..587B}, and the critical density of [CII] and [NII]\,205$\mu$m are nearly the same. The observed [CII]/[NII] ratio is nearly constant ($\sim$4) in the ionized ISM over a large range of electron density, n$_e$. The difference between the predicted and measured [CII]/[NII]205$\mu$m ratio can then be used to estimate the ionized and neutral fractions of [CII] emission. 

\setlength{\unitlength}{1cm}
\begin{figure}
\begin{tabular}{c}
\put(0,0){\includegraphics[width=8.6cm]{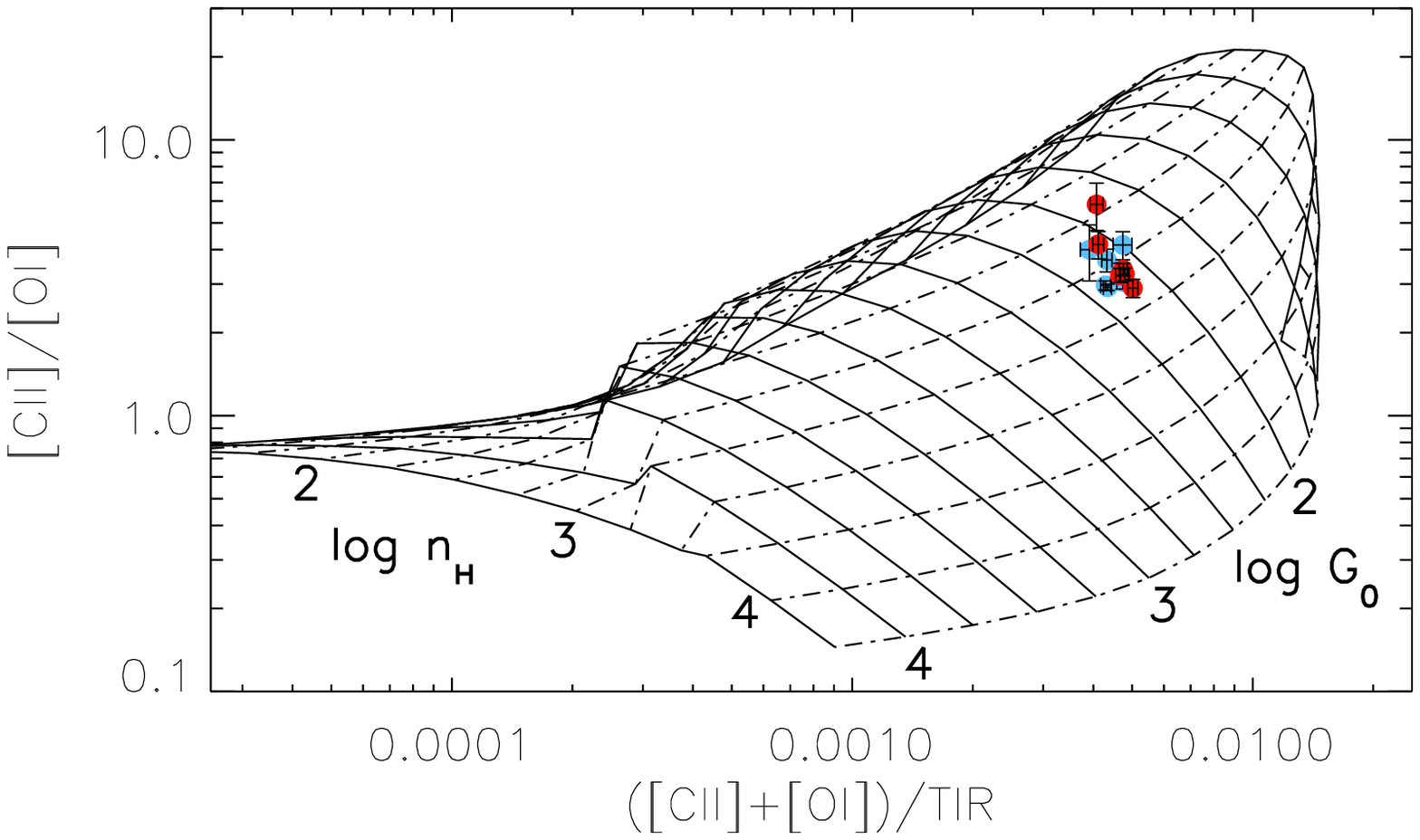}}
\put(1.5,3.2){\includegraphics[width=2.6cm]{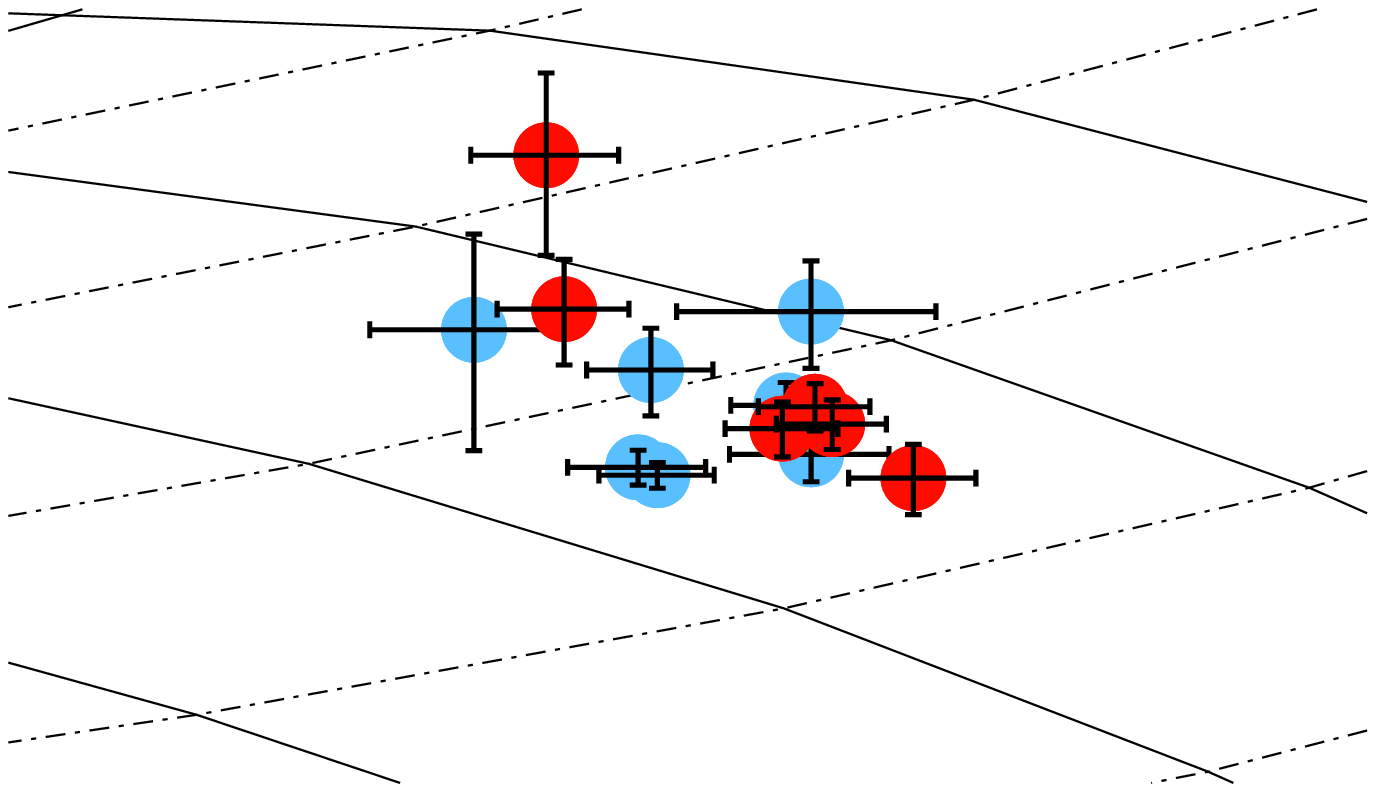}}
\end{tabular}
\caption{Plot of [CII]/[OI] vs. ([CII]+[OI])/TIR to determine the ambient UV radiation field incident on the PDR regions. The [CII] emission has been corrected for a 10\% contribution from the diffuse ionized ISM. The underlying grid is the \citet{1999ApJ...527..795K} PDR model for different densities and stellar radiation field strengths. The inset sub-panel is a zoom-in of the region of interest. Data points in the east/west side of the ring are indicated in blue/red.}
\label{fig:gnplot}
\end{figure}

\setlength{\unitlength}{1cm}
\begin{figure}
\begin{tabular}{c}
\put(0,0){\includegraphics[width=8.6cm]{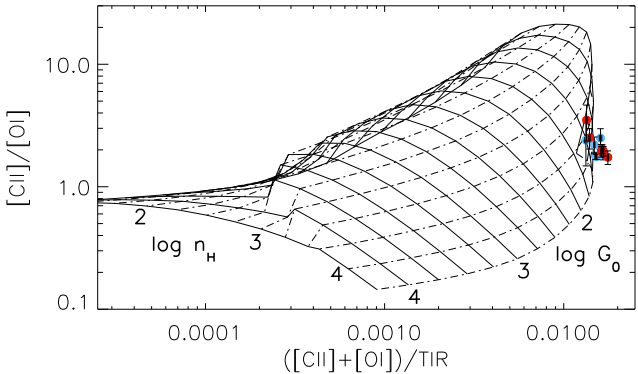}}
\end{tabular}
\caption{Same as Fig. \ref{fig:gnplot}, except with a further assumption of 20\% TIR, and 60\% [CII] contribution from PDRs illuminated by a high radiation field intensity, with the rest originating from more diffuse neutral ISM components.}
\label{fig:gnplot_fpdr}
\end{figure}

Since the [NII]205$\mu$m line is not significantly detected in the NGC\ 4736 ring, we follow an analysis similar to the one discussed in \citet{2012ApJ...747...81C} and use the [CII]/[NII]122$\mu$m ratio to determine the ionized [CII] emission. The [CII]/[NII]122$\mu$m ratio is dependent on electron density in the range 1-100 cm$^{-3}$.  While the [SIII]18.7/33.5$\mu$m ratio is available from our IRS maps, the ratio [SIII]18.7/33.5 is sensitive to much higher electron densities in the range $10^{2-5}$\,cm$^{-3}$, and the [SIII] ratio (0.56) is consistent with gas in the low density limit.

The [NII]122/205 ratio is sensitive to electron densities of 0.1-100 cm$^{-3}$. Since the [NII] 205$\mu$m line is not detected in the ring, the observed [NII]122$\mu$m signal-to-noise ratio in the ring ($5-10$) can be used as a proxy for the upper limit of the [NII]122/205 ratio. From \citet{2012ApJ...751..144B} we know that [NII]122$\mu$m/[NII]205$\mu$m increases with increasing electron density, with the ratio reaching 10 at n$_{e}$ = 100\,cm$^{-3}$. Thus, in the ring we are dealing with an n$_{e}$ lower than 100\,cm$^{-3}$, consistent with our Spitzer/IRS measurements. At an n$_{e}$ of 100\,cm$^{-3}$ the [CII]/[NII]122$\mu$m ratio from the diffuse ionized component of the ISM would be 1.24 (this is an upper limit).  All additional [CII] emission should be from the neutral component. In the region of the ring for which we have [NII]122$\mu$m emission maps, we find [CII]/[NII]122$\mu$m ratios of 8-16, with the value increasing away from the highest intensity regions. These values correspond to 85 to 92\% contributions of the neutral phase to the [CII] emission. At lower electron density the [CII] fraction in the neutral phase rises. We therefore assume that 90\% of the measured [CII] emission in the ring is from the neutral phase.

In Fig.\,\ref{fig:gnplot} the (diffuse ionized emission corrected) [CII]/[OI] and ([CII]+[OI])/TIR ratios are overlaid on a PDR model from \citet[][hereafter K99]{1999ApJ...527..795K}. As described in Sect. \ref{sect:data}, the [CII], [OI], and TIR fluxes were extracted in square apertures of 12\arcsec\, by 12\arcsec\, size, covering the circumnuclear ring. No difference is seen between the two (Fig. \ref{fig:gnplot}, red/blue points) sides of the ring. Based on these [CII]/[OI] and ([CII]+[OI])/TIR ratios the radiation field G$_0$, a dimensionless variable in multiples of the radiation field in the Solar neighborhood, in the PDR regions of the circumnuclear ring is found to have a uniform value of $\sim$200, and the gas density, n$_H$, in the PDR regions to be $\sim$\,10$^{3}$\,cm$^{-1}$. 

We can compare our G$_0$ derived above from the FIR emission lines with the average $<$U$>$ obtained from modeling the dust emission. $<$U$>$ is a measure of the interstellar radiation field in units of the solar neighborhood \citep{2007ApJ...657..810D}. For NGC\, 4736, $<$U$>$ is roughly 10 (Aniano et al., priv. comm.), about a factor of 20 lower than the G$_0$ inferred above. This large mis-match between the radiation field intensity derived from the gas and dust has been noted in other studies as well \citep{2012ApJ...747...81C}, and can be understood by noting that the PDR fraction, the fraction of the dust illuminated by the high intensity radiation field, can be much less than unity.  Aniano et al. find a f$_{PDR}$ (the fraction of dust mass illuminated by G$_{0}$ greater than $\sim$110) in the ring of order 20\%. If we apply this fraction to the TIR, and assume only 60\% of the neutral [CII] emission is from dense PDRs illuminated by the high radiation field intensity, we can reconcile the two estimates (see Fig. \ref{fig:gnplot_fpdr}).

\section{Heating-cooling balance} 
In the `pearls-on-a-string' model, the heating in the ring as a function of azimuth is greatest close to the inflow points, followed by a slow decay as the stars age and the ring rotates. To see if the observed cooling in the ring follows this pattern, we compare the dust continuum flux and the far-IR line emission, against the expected available stellar heating.

\subsection{Energy input into the ISM}
\begin{figure} 
\includegraphics[width=8.6cm]{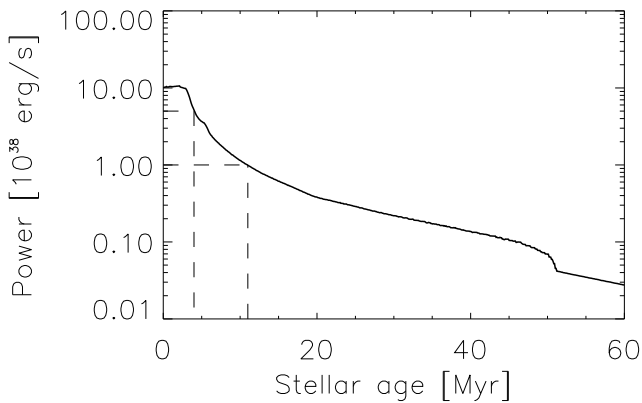}
\caption{Integrated FUV flux between 6-13.6\,eV in the case of an instantaneous starburst, rescaled to 1M$_{\odot}$. Within 4\,Myr the energy has halved, and within 11\,Myr the energy has dropped by an order of magnitude. (Based on Starburst99 v6.0.3. models)}
\label{fig:LW_curve}
\end{figure}

By making several reasonable assumptions, the available energy for dust and gas heating from the newly formed stars in the ring can be derived from the measured SFRD. An instantaneous starburst was modeled using Starburst99 \citep{1999ApJS..123....3L,2010ApJS..189..309L}, assuming a standard Kroupa IMF and solar metallicity. The output synthetic line spectrum was integrated between 912\,\AA{} and 1120\,\AA{} (6-13.6eV), to obtain the photo-electric heating energy as a function of time. The resulting available heating power per M$_{\odot}$ is shown in Fig. \ref{fig:LW_curve}. As can be seen, the power in the 6\,eV to 13.6\,eV range of the integrated stellar population spectrum is strongly dependent on time. The flux plateaus in the first 3\,Myr, but within 11\,Myr, it has dropped by an order of magnitude.  

Instead of a single instantaneous starburst, physically the star formation will occur over a longer time interval. Like a Ferris wheel, the ring can be thought of as having discrete volume elements (`cabins'), scooping up gas over a fixed time interval as it enters onto the ring. In this model, each volume element will have a star formation event lasting a fixed fraction of the rotation time, with the mean age of the event depending on the distance from the inflow point. By tracing back the observed star formation under this premise, we can recreate the heating flux throughout the ring. 
As a reminder: the average SFRD measured in the ring was 0.065\,M$_{\odot}$/yr/kpc$^2$.  To be able to compare the available heating to the observed TIR emission later on, we multiply by an average 24$\mu$m/UV flux ratio of 0.33, based on our observations, as a proxy of how much FUV flux is heating the dust, rather than immediately escaping the galaxy. This ratio is close to the measured dust/stellar flux ratio of 0.25 by \citet{2011ApJ...738...89S}.

\begin{figure*} 
\begin{tabular}{c c}
\includegraphics[width=8.6cm]{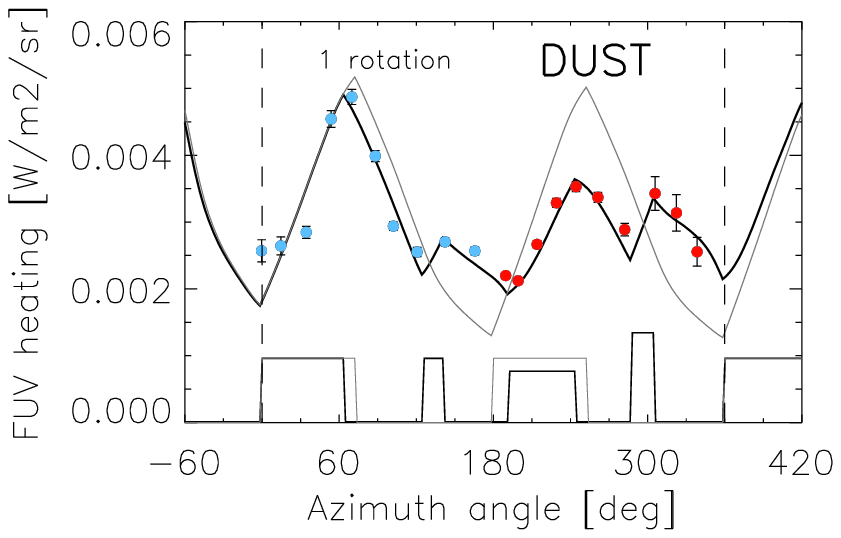} &
\includegraphics[width=8.6cm]{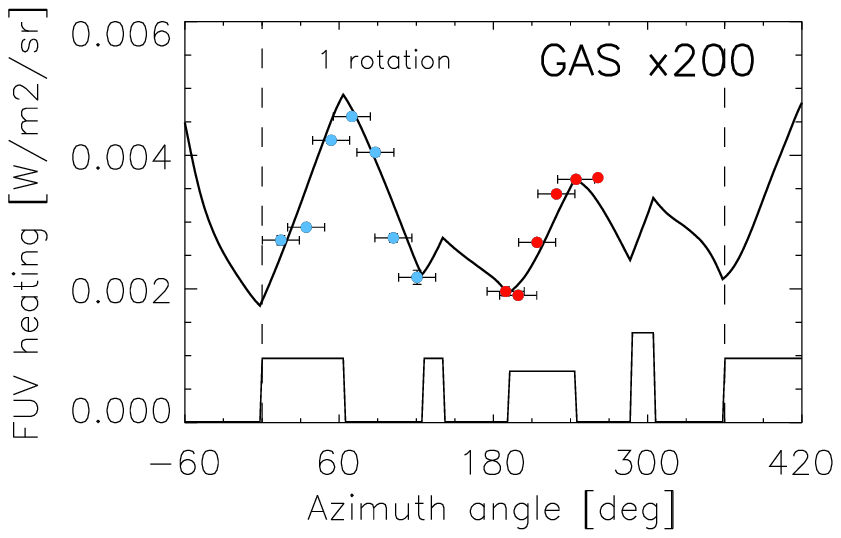} \\
\end{tabular}
\caption{{\it Left:} Available FUV heating flux (black saw-tooth line) as a function of azimuth, derived from an proposed  star formation history. The proposed star formation history is indicated at the bottom of the panel (black box function). The grey lines (underlying the black) are an alternative example of a star formation history and FUV heating flux. The observed FIR (dust) flux is shown in blue/red (east/west side of the ring). Details in the text. {\it Right:} Similar, except now for the [CII]+[OI] (gas) emission lines. The FoV of the Herschel spectra does not cover the full ring, thus for several apertures the gas information is lacking.}
\label{fig:heatcool}
\end{figure*}

\subsection{Comparison to dust}
A SFR history (for one full 25\,Myr orbit) is postulated where there are two starbursts of 5\,Myr duration, 12.5\,Myrs apart, with a burst SFR intensity of 0.01\,M$_{\odot}$/yr (thus a total of 5$\times$10$^4$M$_{\odot}$ stellar mass per starburst is created). The heating curve is modeled by convolving the instantaneous starburst, shown in Fig. \ref{fig:LW_curve}, with this proposed (box-)function of SFR history to obtain the instantaneous starburst heating as a function of time/position throughout the ring (left panel Fig. \ref{fig:heatcool}, grey saw-tooth curve). This heating curve is compared with the TIR continuum sampled throughout the ring (colored points). We may assume an instantaneous response from the dust to the heating, since the timescales here are of order $\sim$Myr, while dust responds to heating on order 10$^4$yr, regardless of dust geometry (B. Draine, priv. comm.). In Fig. \ref{fig:heatcool} the correspondence between the {\it measured} TIR continuum and the {\it modeled} heating is very good. However, it is clear that the heating power falls too steeply around the ring, and that the second peak is significantly lower than the first. A much better match (black curve) can be made by shortening both initial bursts, lowering of the second burst to 0.008\,M$_{\odot}$/yr, and adding two short bursts of star formation at $\sim$8.75-10Myr and 20-21.25Myr. The inclusion of the latter is a divergence from the `pearls-on-a-string' scenario.

\subsection{Comparison to PDR gas}\label{sect:gascool}
\begin{figure}
\includegraphics[width=8.6cm]{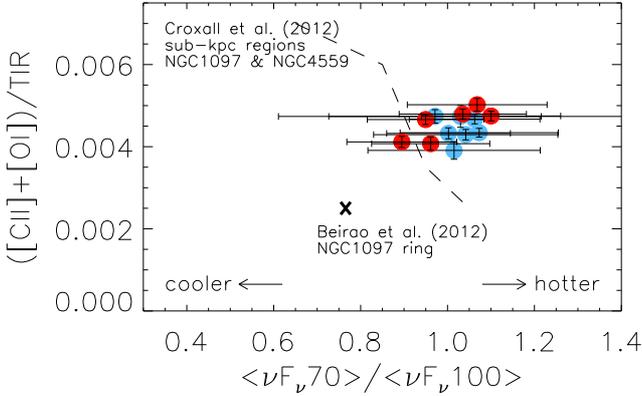}
\caption{Gas heating efficiency proxy, ([CII] + [OI])/TIR, as a function of the dust temperature (blue (east) and red (west) dots). For comparison, the results from  \citet{2012ApJ...751..144B} in the circumnuclear star forming ring of NGC\,1097 (`x'), and \citet{2012ApJ...747...81C} in sub-kpc regions in NGC\,1097 and NGC\,4559 (dashed line) are indicated. The [CII] has been corrected for its 10\% diffuse ionized component.}
\label{fig:heatciioi}
\end{figure}

To compare the response of the PDR gas to the modeled heating we need an observational measure of the photoelectric heating efficiency. This is the ratio of photon energy going into gas heating to the total available energy. This ratio, typically taken as ([CII] + [OI])/TIR, is equal to the ratio of gas cooling line flux to TIR continuum flux. In Fig. \ref{fig:heatciioi} this gas to dust ratio is plotted against, the ratio $\langle\nu\,S_{\nu}\rangle_{70}/\langle\nu\,S_{\nu}\rangle_{100}$, which is used as a proxy for the dust temperature. From Fig. \ref{fig:heatciioi} it can be seen that the photoelectric heating efficiency shows very little spread over the circumnuclear ring, with values of 0.004-0.006, about 0.5\%.  

In the right panel of Fig. \ref{fig:heatcool}, this fixed 0.5\% scaling has been applied to the [CII]+[OI] flux in each region. It is immediately clear that the spread in the emission line flux is due to the FUV heating variation. The two cooling lines again follow the modeled FUV heating quite closely, although we lack the data points away from the inflow points to confirm the secondary star forming events, due to limited Herschel spectral coverage.

The photoelectric heating efficiencies are comparable to the values seen for other kpc-sized star forming regions, as for example in \citet{2012ApJ...747...81C} or \citet{2013ApJ...776...65P}. These studies find values of $\sim$0.5\% at lower dust temperatures, with a significant drop of heating efficiency at higher dust temperatures. However, \citet{2012A&A...548A..91L} and \citet{2014ApJ...787...16P} do not find a reduction in the gas heating efficiency at higher dust temperatures in a star forming region of the LMC, and the disk of Centaurus A, respectively. The NGC\,4736 ring does, however, have a higher photoelectric heating efficiency, at only marginally higher dust temperature, than starburst ring in NGC\,1097 ring \citep{2012ApJ...751..144B}.

\section{Implications for ring star formation theory}
We started this investigation to study if, following the `pearls-on-a-string' paradigm for circumnuclear star forming rings, all star formation takes place at the inflow points of the ring. In this scenario, as the system rotates, the subsequent star forming events should create a clear azimuthal age gradient \citep{Boker2008}. 

Our simple modeling of the PDR dust and gas heating as function of azimuth requires secondary star formation events to be consistent with the observed cooling flux. While the exact strength, timing, and duration of these secondary events in the NGC\,4736 ring is model dependent, it is clear that their absence leads to a too large contrast in heating/cooling flux around the ring. Since the FUV heating power available from young stars diminishes by an order of magnitude in 11\,Myr, it seems unlikely that the drop be compensated for by stellar 'after-glow'. 

Are these `secondary' star forming events truly random, taking place from left-over gas from the `primary' event? Or, are both events part of the `popcorn' scenario \citep{Elmegreen1994}, where star formation occurs anywhere in the ring? Given the relative strength of the events at both inflow points we consider this unlikely. This seems consistent with the fact that a number of galaxies show strong star formation events at the inflow points of their circumnuclear rings \citep{2001MNRAS.323..663R,2005ApJ...633L..25A,Boker2008,2008ApJS..174..337M,Tessel2,Tessel3}. 

A combination of `pearls-on-a-string' and `popcorn' seems like a reasonable scenario to explain the star formation in the ring of NGC\ 4736. We know star formation is not a 100\% efficient process. Star formation efficiencies (SFE) are usually estimated to be of order a few percent. Although there may be an indication that the SFE in the center of NGC\,4736 is actually higher (10\%, K. Sandstrom, priv. comm.), even then a significant portion of gas remains. Therefore, we can ask if the remaining gas has enough chance to cool and is dense enough to collapse again.

The total gas mass surface density in the ring remains $\sim$50-80\,M$_{\odot}$/pc$^2$ away from the inflow points (Fig. \ref{fig:gas}). This is higher than the $\sim$8\,M$_{\odot}$/pc$^2$ star formation threshold noticed in \citet{2008AJ....136.2846B}. Additionally, while the temporal distance between the inflow points and these modeled secondary bursts is short (a few Myr), it is comparable to free fall times of giant molecular clouds. 

Can we expect this hybrid star formation scenario in all circumnuclear rings? The life times of circumnuclear rings are of order 1-2\,Gyr. Inflow rates of gas onto the ring of 2\,M$_{\odot}$/yr are a reasonable upper limit \citep{Tessel,Tessel2}, although values of 0.1-0.25\,M$_{\odot}$/yr are more likely. So, even at extreme SFEs and low gas inflow rates, the gas mass in a ring will see an increase of about 10$^8$\,M$_{\odot}$ (0.1\,M$_{\odot}$/yr$\times$1\,Gyr) over the life time of the ring. Given the observed range of radii and widths found for circumnuclear rings (40-1000\,pc radii, $\sim$r/2 widths, \citet{Comeron2010}), this translates easily to gas mass surface densities similar to those found in NGC\ 4736 (10-10$^2$\,M$_{\odot}$/pc$^2$). Thus, the prevalence of `popcorn' star formation events should increase with the lifetime of the circumnuclear ring, even if starbursts, `pearls-on-a-string', continue to be at the inflow points.

\section{Summary}
New Herschel FIR observations of the circumnuclear ring in NGC\,4736 have been combined with archival observations of the star formation, and atomic and molecular gas distributions. This combination of data sets is used to characterize the mode of star formation in the ring. Due to the circular orbits of the ring itself, gas and dust will only experience large-scale shocks and compression at the inflow points, the locations where material transitions onto the ring. Star formation events, therefore, should be uniquely found close to the inflow points. We have investigated the heating and cooling of the interstellar medium in the ring of NGC\,4736, under this assumption. The stellar FUV flux available for heating was derived from measurements of the star formation rate, geometry-based estimates of time scales, and standard assumptions on the initial mass function (IMF). The FIR continuum and gas emission line fluxes were measured from Herschel PACS and archival Spitzer observations, and used to estimate the heating and cooling of the dust and gas around the ring.

We find that, while heating from a significant burst of star formation at the inflow points can explain the gas and dust cooling there, an additional star formation component is required at other positions in the ring. This additional component is smaller than the star formation events at the inflow points, and most likely arises from the general increase of gas density in the ring over its lifetime. Therefore, in the NGC\ 4736 ring, it is a combination of the `popcorn' and `pearls-on-a-string` models that best explains the data.

\begin{acknowledgements}
This project was started during a '11/'12 IPAC visiting graduate fellowship awarded to TvdL. TvdL would like to thank IPAC, especially P. Appleton and L. Armus, for hosting her. 
\end{acknowledgements}

\bibliographystyle{aa}
\bibliography{papers.bib}
\end{document}